\documentclass[a4,fleqn,usenatbib]{mnras}

\usepackage[T1]{fontenc}
\usepackage{ae,aecompl}

\usepackage{graphicx}	
\usepackage{amsmath}	
\usepackage{amssymb}	
\usepackage{breakurl}

\title[MAGIC detection of H1722+119]{Investigating the peculiar emission from the new VHE gamma-ray source H1722+119}

\author[M.~L.~Ahnen~et.~al.]{
M.~L.~Ahnen$^{1}$,
S.~Ansoldi$^{2}$,
L.~A.~Antonelli$^{3}$,
P.~Antoranz$^{4}$,
A.~Babic$^{5}$,
B.~Banerjee$^{6}$,\newauthor
P.~Bangale$^{7}$,
U.~Barres de Almeida$^{7,24}$,
J.~A.~Barrio$^{8}$
% ...
(The MAGIC Collaboration),
F.~D'Ammando$^{31,32}$(for the {\it Fermi}-LAT
Collaboration),\newauthor
A.~Berdyugin$^{33}$,
T.~Hovatta$^{34,35}$,
W.~Max-Moerbeck$^{36}$,
C.~M.~Raiteri$^{37}$,\newauthor
A.~C.~S.~Readhead$^{35}$,
R.~Reinthal$^{33}$,
J.~L.~Richards$^{38}$,
F.~Verrecchia$^{39,40}$
and M.~Villata$^{36}$\newauthor
(Affiliations can be found after the references)
}

\date{Accepted XXX. Received YYY; in original form ZZZ}

\pubyear{2016}

\begin{document}
\label{firstpage}
\pagerange{\pageref{firstpage}--\pageref{lastpage}}
\maketitle

\begin{abstract}
The Major Atmospheric Gamma-ray Imaging Cherenkov (MAGIC) telescopes observed the BL Lac object H1722+119 (redshift unknown) for six consecutive nights between 2013 May 17 and 22, for a total of 12.5\,h. The observations were triggered by high activity in the optical band measured by the KVA (Kungliga Vetenskapsakademien) telescope. The source was for the first time detected in the very high energy (VHE, $E > 100$\,GeV) $\gamma$-ray band with a statistical significance of 5.9\,$\sigma$. The integral flux above 150\,GeV is estimated to be $(2.0\pm 0.5)$ per cent of the Crab Nebula flux. We used contemporaneous high energy  (HE, 100\,MeV $ < E < 100$\,GeV) $\gamma$-ray observations from {\it Fermi}-LAT (Large Area Telescope) to estimate the redshift of the source. Within the framework of the current extragalactic background light models, we estimate the redshift to be $z = 0.34 \pm 0.15$. Additionally, we used contemporaneous X-ray to radio data collected by the instruments on board the {\it Swift} satellite, the KVA, and the OVRO (Owens Valley Radio Observatory) telescope to study multifrequency characteristics of the source. We found no significant temporal variability of the flux in the HE and VHE bands. The flux in the optical and radio wavebands, on the other hand, did vary with different patterns. The spectral energy distribution (SED) of H1722+119 shows surprising behaviour in the $\sim 3\times10^{14} - 10^{18}$\,Hz frequency range. It can be modelled using an inhomogeneous helical jet synchrotron self-Compton model. 
\end{abstract}




\section{Introduction}
\label{Introduction}

Active galactic nuclei (AGN) are the most luminous persistent sources of electromagnetic radiation in the Universe. They are compact regions in the centres of galaxies formed around supermassive black holes (SMBHs) that actively accrete matter. Approximately 10 per cent of AGN eject matter through relativistic jets \citep{Dunlop03}. AGN whose jets are oriented close to the line of sight to the Earth are referred to as blazars. Jets are sources of electromagnetic radiation of all wavelengths from radio to $\gamma$-rays. They can extend to Mpc distances from the nucleus and can be brighter than the rest of the galaxy. The spectral energy distribution (SED) of blazars is characterised by two broad peaks: a ``low-energy" peak in the optical to X-ray band and a ``high-energy" peak in the X-ray to $\gamma$-ray band. Their emission is characterised by the strong and variable linear polarisation in the optical and radio bands, and high variability in flux across the entire electromagnetic spectrum. Blazars can be divided into two classes according to the characteristics of their emission: flat spectrum radio quasars (FSRQs) and BL Lacertae objects (BL Lacs). FSRQs are known to have prominent broad and narrow optical emission lines, in addition to strong optical and X-ray continuum emission. A quite common feature is the so-called blue bump in the optical-UV band associated with the emission from the accretion disc. BL Lacs, on the other hand exhibit very weak optical emission lines if any at all. 
The low-energy peak in SED of FSRQs is believed to be a combination of synchrotron radiation of electrons in the jet and thermal emission from the broad line region (BLR), dust torus and accretion disc, while usually in the case of BL Lacs, all the emission is attributed to synchrotron emission. According to leptonic scenarios, the second peak is a result of inverse Compton (IC) scattering of lower-energy photons (so-called seed photons) on relativistic electrons within the jet, while hadronic scenarios assume protons in the jet are accelerated even to energies $\ga 10^{19}$\,eV and significantly contribute to the emission either through proton-synchrotron emission in relatively strong magnetic fields, or photo-pion production \citep{Boettcher2013}. 
\citet{Ghisellini2010} showed that the broad-band SEDs of $\gamma$-ray blazars can be, on average, well described by a simple one-zone leptonic model including synchrotron and IC emission components, with the addition of possible external contributions from e.g. the accretion disc or host galaxy emission. 

H1722+119 was first observed in the 1970s as a part of the {\it Uhuru} X-ray sky survey and the {\it HEAO 1} Large Area Sky Survey (LASS). The resulting X-ray source catalogues (\citet{FormanUhuru}; source name: 4U1722+11, and \citet{HEAOWood}; source name: 1H1720+117) only identify it as an X-ray source. Almost twenty years later, \citet{Griffiths} and \citet{Brissenden} independently, classified it as a BL Lac object. Furthermore, H1722+119 was classified as an intermediate-energy-peaked BL Lac with the low-energy peak at $\nu_{s} = 6.3\times10^{15}$\,Hz \citep{NieppolaCatalog}. \citet{Brissenden} measured a very high level of linear polarization in the optical band, reaching $17.6 \pm 1.0$ per cent in the B band.
H1722+119 has been observed in the radio band by the OVRO (Owens Valley Radio Observatory) since 2007 \citep{OVRO}\footnote{\href{http://www.astro.caltech.edu/ovroblazars}{www.astro.caltech.edu/ovroblazars}}.
H1722+119 is included in the third {\em Fermi}-LAT (Large Area Telescope) catalogue \citep[3FGL;][]{Acero2015} as a counterpart of the $\gamma$-ray source 3FGL\,J1725.0+1152, with photon index $\Gamma = 1.89\pm0.05$ and 0.1--100\,GeV flux of $(2.7\pm0.3)\times10^{-8}$\,ph\,cm$^{-2}$\,s$^{-1}$.
Interesting studies, in which radio and high energy  (HE, 100\,MeV $ < E < 100$\,GeV) $\gamma$-ray properties of blazars and their correlations are discussed, were presented in \citet{MOJAVEProgram} and \citet{Linford1}. Both works used the LAT on board the {\it Fermi Gamma-ray Space Telescope} for HE $\gamma$-ray and the Very Long Baseline Array for radio observations. \citet{Linford1} (source name: F17250+1151) describe H1722+119 as a compact source with a short jet.
Infrared (IR) observations of H1722+119 were performed as part of the Two Micron All-Sky Survey \citep{2MASS}.
H1722+119 was listed as a candidate TeV blazar in \citet{Costamante_Ghisellini} based on its X-ray and radio properties.
Observations in the very high energy (VHE, $E > 100$\,GeV) $\gamma$-ray band were first reported in \citet[][source name: RX J1725.0+1152]{MAGICStacked}. The MAGIC (Major Atmospheric Gamma-ray Imaging Cherenkov) telescope observed H1722+119 between 2005 and 2009 together with 20 other BL Lac candidates. H1722+119 was selected for this campaign based on its X-ray properties from \citet{Donato}.
The stacked sample of observed blazars showed a signal above 100\,GeV with a significance of 4.9\,$\sigma$, but the analysis of H1722+119 data alone resulted in 1.4\,$\sigma$ after 32\,h of observation, with an upper limit (UL) of flux above 140\,GeV of $1.3\times 10^{-11}\,{\rm ph\,cm^{-2}\,s^{-1}}$.
\parskip=0pt

Although \citet{Brissenden} reported a featureless optical spectrum, \citet{Griffiths} estimated the redshift of the host galaxy based on an absorption feature to be $z = 0.018$. However, this result was not confirmed by other optical observations \citep[e.g.][]{VeronCetty, Falomo1, Falomo2}. \citet{Sbarufatti} observed H1722+119 with the ESO Very Large Telescope and detected no intrinsic features in the optical spectra. They derived a lower limit (LL) of $z > 0.17$. \citet{Landoni2014} used the spectrograph X-Shooter at the European Southern Observatory Very Large Telescope to set a LL on the redshift at 0.35. They detected no intrinsic or intervening spectral lines, ascribing this to extreme optical beaming, setting the ratio of beamed to thermal emission at $\ge400$. Farina et al. (2013, priv. comm.) observed H1722+119 with the NOTCam of the Nordic Optical Telescope in the H-band in 2013 and were unable to detect the host galaxy. Applying the imaging redshift technique proposed by \citet{Sbarufatti2005}, they set a LL on redshift at 0.4.

In this paper we report the first detection of VHE $\gamma$-ray emission from H1722+119, by the MAGIC telescopes \citep{MAGICAtel}, and study the multifrequency characteristics of H1722+119 in that period.
Emission from blazars is quite variable in time, and optical high states are often used to trigger MAGIC observations. H1722+119 joins quite a long list of blazars, whose VHE $\gamma$-ray signal was detected following an optical high state \citep[see e.g.][]{Albert2006_Mrk180,Albert2007_1ES1011,Albert2008_HBL,Anderhub2009_S50716,Aleksic2012_B32247,Aleksic2012_1ES1215,Aleksic2014_1ES1727,Aleksic2015_1ES0806}. 
In Section \ref{Observations} we present the instruments used in this work and their respective results. Section \ref{MWL} is reserved for study of multifrequency characteristics. We summarise our findings in Section \ref{summary}. 
Throughout the paper, we assume standard $\Lambda$CDM cosmology \citep{Komatsu2011}.

\section{Observations and Data Analysis}
\label{Observations}

\subsection{MAGIC}
\label{MAGIC}
The MAGIC telescopes are located at the Observatorio del Roque de los Muchachos in the Canary Island of La Palma, Spain ($28^{\circ}45^{\prime}$ north, $18^{\circ}54^{\prime}$ west), at 2200\,m above sea level. Two 17\,m diameter Imaging Atmospheric Cherenkov Telescopes are optimised for observations of $\gamma$-rays of energies above 50\,GeV. 
A detailed overview of the MAGIC experiment and the telescope performance is given in \citet{MAGICperform_1, MAGICperform_2}.

The results reported here are based on the observations performed during six nights between 2013 May 17 and 22, triggered by the high optical state detected by the KVA (Kungliga Vetenskaps Akademientelescope) 
(see Section \ref{KVA}). Between the first (2005--2009) and the second (2013) observation campaign the MAGIC system underwent a series of major upgrades \citep[see][]{Perform_First,Perform_Split,Perform_2GHz,Perform_Stereo,Perform_DRS4,MAGICperform_1,MAGICperform_2}. 
The current instrument is roughly twice as sensitive in the $100-200$\,GeV energy range compared to the single MAGIC I telescope in operation in 2009 \citep{MAGICperform_2}. 

MAGIC usually observes sources in the so called {\it wobble} mode \citep{wobble,MAGICperform_2}. 
H1722+119 was observed in four false-source positions for a total of 12.5\,h during 2013. After quality selection based on the stability of the event rates, the effective time amounted to 12.0\,h. 
Observations were performed at zenith angles between $16^{\circ}$ and $37^{\circ}$.

\begin{figure}
\centering
\includegraphics[width=0.5\textwidth]{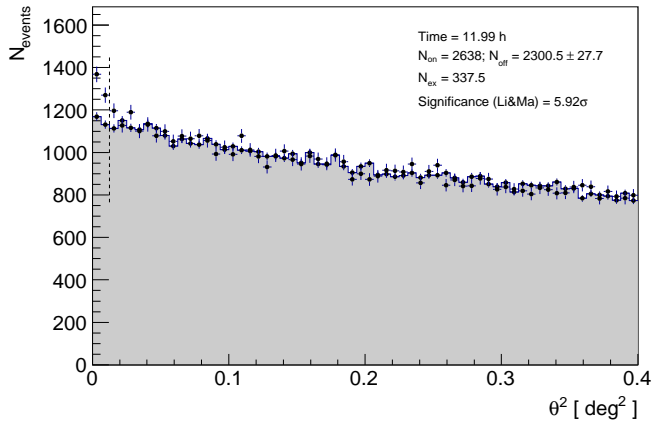}
\caption{$\theta^2$ distribution of signal (black points) and background (grey area) events. $N_{\rm on}$ is the number of all events with $\theta^2<0.0125\,\rm deg^2$ (vertical dashed line) with respect to the source position in the camera, and $N_{\rm off}$ the normalised number of all events from the same region around OFF-source position.}
\label{fig:Theta2}
\end{figure}

\begin{figure}
\centering
\includegraphics[width=0.5\textwidth]{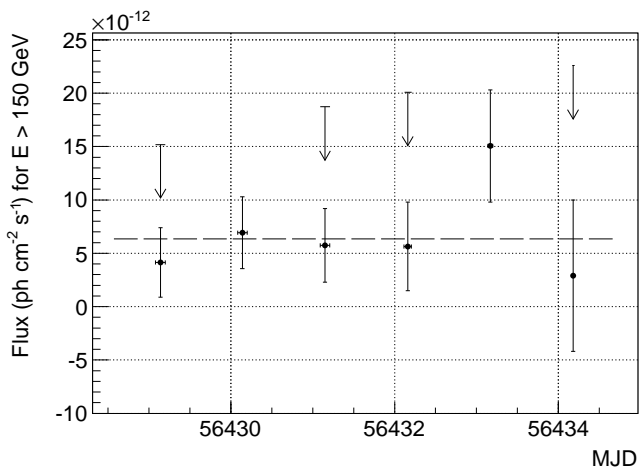}
\caption{MAGIC nightly light curve for energies above 150\,GeV. Horizontal error bars represent the duration of each observation. Vertical arrows represent ULs for points whose relative error of the excess is $> 0.5$. The horizontal dashed line is a constant flux fit with parameters stated in the text.}
\label{fig:MAGIC_LC}
\end{figure}

The data were analysed within the \texttt{MARS} (MAGIC Analysis and Reconstruction Software) analysis framework \citep{Lombardi,MARS}. 
The VHE $\gamma$-ray signal is estimated using the distribution of the squared angular distance ($\theta^2$) between the reconstructed and nominal (ON-source) source positions in the camera coordinates for each event. Background is estimated in the same manner with respect to the OFF-source position. Usually three OFF-source positions are chosen at the same distance from the camera centre as the ON-source position and rotated by $90^\circ$ each.
$N_{\rm on}$ is the number of events originating within the source region ($\theta^2<0.0125\,\rm deg^2$), and $N_{\rm off}$ the normalised number of all events from the same region around OFF-source positions. An excess of 337.5 events above 60\,GeV with respect to the background was detected, yielding a signal significance of 5.9\,$\sigma$ using Eq. 17 of \citet{LiMa}. The $\theta^2$ distribution is shown in Fig.~\ref{fig:Theta2}.
The light curve of the integral VHE $\gamma$-ray flux above 150\,GeV is shown in Fig.~\ref{fig:MAGIC_LC}. There is no evidence of flux variability on a night-by-night basis. A fit with a constant flux of $(6.3 \pm 1.6) \times 10^{-12}\,{\rm ph\,cm^{-2}\,s^{-1}}$ resulted in $\chi ^2 / NDF = 3.5/5$, where $NDF$ stands for number of degrees of freedom. This is equivalent to $(2.0 \pm 0.5)$ per cent of the Crab Nebula VHE $\gamma$-ray flux. The measured flux is consistent with the UL set by the MAGIC observations of H1722+119 from the previous campaign \citep{MAGICStacked}.
The differential energy spectrum was reconstructed using the Forward Unfolding algorithm presented in \citet{ForwardUnfold}. It can be described by a simple power law function $dN/dE = f_0 (E/E_0)^{-\Gamma}$ with normalization 
$f_0 = (4.3 \pm 0.9_{\rm stat} \pm 0.9_{\rm syst}) \times \rm 10^{-11}\,ph\,cm^{-2}\,s^{-1}\,TeV^{-1}$ 
and a photon index 
$\Gamma = 3.3 \pm 0.3_{\rm stat} \pm 0.2_{\rm syst}$, 
at normalization energy $E_0 = 200$\,GeV. The fit resulted in $(\chi ^2 / NDF = 3.6/8)$. The systematic uncertainty of the photon index was estimated to be $\pm 0.2$, using Eq. 3 of \citet{MAGICperform_2}. We estimate the error on the $f_0$ to be 20 per cent, which does not include the energy scale uncertainty estimated to be 17 per cent.

\subsection{{\em Fermi}-LAT}
\label{Fermi}
The {\it Fermi}-LAT is a pair-conversion telescope operating from 20 MeV to $>$ 300 GeV. Details about the {\it Fermi}-LAT are given in \citet{atwood09}. The LAT data reported in this paper were collected from 2013 January 1 (MJD 56293) to December 31 (MJD 56657). During this time, the {\it Fermi} observatory operated almost entirely in survey mode. The analysis was performed with the \texttt{ScienceTools} software package version v9r32p5. Only Pass 7 reprocessed events belonging to the `Source' class were used. The time intervals when the rocking angle of the LAT was greater than 52$^{\circ}$ were rejected. In addition, a cut on the zenith angle ($< 100^{\circ}$) was applied to reduce contamination from the Earth limb $\gamma$ rays, which are produced by cosmic rays interacting with the upper atmosphere. The spectral analysis was performed with the instrument response function \texttt{P7REP\_SOURCE\_V15} using an unbinned maximum-likelihood method implemented in the tool \texttt{gtlike} in the $0.1-100$\,GeV energy range. Isotropic (\texttt{iso\_source\_v05.txt}) and Galactic diffuse emission (\texttt{gll\_iem\_v05\_rev1.fit}) components were used to model the background\footnote{\href{http://fermi.gsfc.nasa.gov/ssc/data/access/lat/BackgroundModels.html}{fermi.gsfc.nasa.gov/ssc/data/access/lat/BackgroundModels.html}}. The normalizations of both components were allowed to vary freely during the spectral fitting. 

We analysed a region of interest of $10^{\circ}$ radius centred at the location of H1722$+$119. We evaluated the significance of the $\gamma$-ray signal from the source by means of the maximum-likelihood test statistic TS = 2 (log$L_1$ - log$L_0$), where $L$ is the likelihood of the data given the model with ($L_1$) or without ($L_0$) a point source at the position of H1722$+$119 \citep[e.g.][]{mattox96}. The source model used in \texttt{gtlike} includes all sources from the 3FGL catalogue that fall within $15^{\circ}$ of the source. A first maximum-likelihood analysis was performed to remove from the model the sources having TS $<$ 10 and/or the predicted number of counts based on the fitted model $N_{\rm pred} < 3 $. A second maximum-likelihood analysis was performed on the updated source model. In the fitting procedure, the normalization factors and the photon indices of the sources lying within 10$^{\circ}$ of H1722$+$119 were left as free parameters. For the sources located between 10$^{\circ}$ and 15$^{\circ}$ from our target, we kept the normalization and the photon index fixed to the values from the 3FGL catalogue.

All uncertainties in measured HE $\gamma$-ray flux reported throughout this paper are statistical only. 
The systematic uncertainty in the flux is dominated by the systematic uncertainty in the effective area \citep{ackermann12}, which amounts to 5--10 per cent in the 0.1--100\,GeV energy range, and therefore smaller than the typical statistical uncertainties for this analysis\footnote{\href{http://fermi.gsfc.nasa.gov/ssc/data/analysis/LAT\_caveats.html}{fermi.gsfc.nasa.gov/ssc/data/analysis/LAT\_caveats.html}}.

Integrating over the period 2013 January 1 -- December 31 in the 0.1--100 GeV energy range, using a power law model as in the 3FGL catalogue, the fit yielded a TS = 335, with an average flux of (4.1 $\pm$ 0.7) $\times$10$^{-8}$ ph cm$^{-2}$ s$^{-1}$, and a photon index of $\Gamma$ = 1.99 $\pm$ 0.08 at the decorrelation energy $E_0$ = 3063\,MeV. During the same period, the fit with a log-parabola model, $dN/dE \propto$ $(E/E_{0})^{-\alpha-\beta \, \log(E/E_0)}$, in the 0.1--100 GeV energy range results in TS = 355, with a spectral slope $\alpha$ = 1.99 $\pm$ 0.08, and a curvature parameter $\beta$ = 0.01 $\pm$ 0.01, indicating no significant curvature of the $\gamma$-ray spectrum. The LAT light curve and the photon index evolution with 2-month time bins is shown in Fig.~\ref{fig:MWL}. The 2-month bin width is a trade-off between the significance of the source (TS > 25 in each time bin) and the shortest time-scale to be probed. 
For each time bin, the spectral parameters for H1722$+$119 and for all the sources within 10$^{\circ}$ from the target were left free to vary. No significant increase of the $\gamma$-ray activity was observed by LAT in 2013 May and June -- the time bin contemporaneous to the MAGIC observations. Both the flux and the photon index are well fitted by a constant with the following parameters: flux: $p_0 = (3.43 \pm 0.56)\times 10^{-8}$\,ph\,cm$^{-2}$\,s$^{-1}$ and $\chi^2 / NDF = 3.87/5$, photon index: $\Gamma = 1.96 \pm 0.07$ and $\chi^2 / NDF = 1.68/5$. 
Leaving the photon index free to vary during 2013 May, we obtained TS = 37, with an average flux of (8.5$\pm$3.3)$\times$10$^{-9}$\,ph\,cm$^{-2}$\,s$^{-1}$, and a photon index of $\Gamma = 1.4\pm0.3$. The hint of hardening of the LAT spectrum may be an indication of the shift of the IC peak to higher energies during the MAGIC detection. By considering only the period May 13--26, including the MAGIC observation period, the maximum-likelihood analysis results in a TS = 15, with a flux of $(0.9\pm0.4)\times 10^{-8}$\,ph\,cm$^{-2}$\,s$^{-1}$ (assuming $\Gamma = 1.4$). 
By means of the \texttt{gtsrcprob} tool, we estimated that the highest
energy photon emitted from H1722$+$119 (with probability $>$ 90 per cent
of being associated with the source) was observed on 2013 April 28 (MJD
54610), with an energy of 65.7\,GeV. Another photon with an
energy 51.6\,GeV was observed on 2013 December 31 (MJD 56657).

\subsection{{\em Swift}}
\label{Swift}
The {\it Swift} satellite \citep{gehrels04} performed three observations of H1722$+$119 on 2008 May 31, 2013 January 15 and May 20. The observations were carried out with all three on board instruments: the X-ray Telescope \citep[XRT;][0.2--10.0 keV]{burrows05}, the Ultraviolet/Optical Telescope \citep[UVOT;][170--600 nm]{roming05} and the Burst Alert Telescope \citep[BAT;][15--150 keV]{barthelmy05}. The hard X-ray flux of this source is below the sensitivity of the BAT instrument for the short exposure of these observations (see Table \ref{tab:XRT}), therefore the data from this instrument are not included in this work. Moreover, the source is not present in the {\it Swift}/BAT 70-month hard X-ray catalogue \citep{baumgartner13}.

The XRT data were processed with standard procedures (\texttt{xrtpipeline v0.12.8}), filtering, and screening criteria using the \texttt{HEAsoft}
package (v6.14). The data were collected in photon counting (PC) mode on 2008 May 31 and 2013 January 15, and
windowed timing (WT) mode on 2013 May 20. The source count rate in PC mode was low ($<$ 0.5 counts s$^{-1}$);
thus pile-up correction was not required. Source events were extracted from a
circular region with a radius of 20 pixels (1 pixel = 2.36 arcsec), while
background events were extracted from a circular region with a radius of 50 pixels in PC mode and 20 pixels in WT mode away from the source region and from other bright sources. Ancillary
response files were generated with \texttt{xrtmkarf}, and account for
different extraction regions, vignetting and point-spread function
corrections. We used the spectral redistribution matrices in the Calibration
data base maintained by HEASARC. We fitted the spectrum with an absorbed 
power law using the photoelectric absorption model \texttt{tbabs}
\citep{wilms00}, with a neutral hydrogen column density fixed to its Galactic
value \citep[$N_{\rm H} = 8.88\times10^{20}$\,cm$^{-2}$;][]{kalberla05}. We noted that the 
X-ray flux of the source observed in 2013, i.e. (1--2)$\times10^{-11}$\,erg\,cm$^{-2}$\,s$^{-1}$ (Fig.~\ref{fig:MWL}), is a factor between 3 and 5 higher than the flux in 2008 May 31, $(3.6\pm0.7)\times10^{-12}$\,erg\,cm$^{-2}$\,s$^{-1}$, with no significant spectral change (Table \ref{tab:XRT}).
\begin{table*}
\caption{Log and fitting results of {\em Swift}/XRT observations of H1722$+$119 using a power law model with $N_{\rm H}$ fixed to Galactic
  absorption.}
\begin{center}
\begin{tabular}{ccccc}
\hline
\multicolumn{1}{c}{Observation} &
\multicolumn{1}{c}{Net Exposure Time} &
\multicolumn{1}{c}{Photon index} &
\multicolumn{1}{c}{Flux 2-10 keV} &
\multicolumn{1}{c}{$\chi^2_{\rm\,red}$ $(NDF)$} \\
\multicolumn{1}{c}{Date (MJD)} &
\multicolumn{1}{c}{s} &
\multicolumn{1}{c}{$\Gamma$} &
\multicolumn{1}{c}{$\times$10$^{-12}$ erg cm$^{-2}$ s$^{-1}$} \\
\hline
2008-05-31 (54617) & 1733 & $2.13 \pm 0.22$ & $ 3.6 \pm 0.7$ & 1.257  (11)  \\
2013-01-15 (56307) &  812 & $2.28 \pm 0.15$ & $17.8 \pm 2.3$ & 0.9040 (17)  \\
2013-05-20 (56432) & 1983 & $2.31 \pm 0.08$ & $10.5 \pm 0.7$ & 1.245  (64)  \\
\hline
\end{tabular}
\end{center}
\label{tab:XRT}
\end{table*}
For the observation in 2013 May 20, with the higher statistics, we fit the spectrum also with a log-parabola model, obtaining a spectral slope $\alpha = 2.2 \pm 0.1$, a curvature parameter of $\beta$ = 0.54$^{+0.34}_{-0.31}$, and a $\chi^{2}_{\rm\,red}/NDF = 1.124/63$. The F-test shows an improvement of the fit with respect to the simple power law ($\chi^{2}_{\rm\,red}/NDF = 1.245/64$) with a probability of 99.9 per cent.

UVOT data in the $v$, $b$, $u$, $w1$, $m2$, and $w2$ filters were reduced with the \texttt{HEAsoft} package v6.16 executing the aperture-photometry task \texttt{uvotsource}. 
We extracted the source counts from an aperture of 5$\arcsec$ 
centred on the source, and the background counts from a circle with 10 arcsec
radius in a nearby source-free region. 
Magnitudes were converted into dereddened flux densities
by adopting the extinction value E(B--V) = 0.1497 from \citet{schlafly11}, the mean Galactic extinction curve in \citet{fitzpatrick99} and the magnitude-flux calibrations by \citet{bessel88}. The UVOT flux densities collected in 2013 are reported in Fig.~\ref{fig:MWL}. 
The flux densities observed in 2008 May 31 are a factor of 2 lower with respect to the 2013 observations. 

Checking the source SED (Fig. \ref{fig:SED}), we noticed that a monotonic connection between optical-UV and X-ray spectrum was not possible, which motivated us to analyse possible sources of errors affecting the data. 
An aperture correction procedure was executed for the 2013 May 20 UVOT images in all filters, estimating from field stars the correction to magnitudes extracted with full-width at half maximum (FWHM) radius circular apertures to obtain the counts within the standard apertures (5$\arcsec$ radii). A first photometry of the object with apertures of 2.5$\arcsec$ (the FWHM for all filters) was executed.
Then 6 to 8 non-saturated field stars were selected and photometry was performed for each of them using FWHM radii and standard apertures, and non-contaminated background annular regions of radii 26$\arcsec$ to 33$\arcsec$. The weighted mean among all magnitude differences obtained with the two different apertures was calculated and finally subtracted from the object magnitude at FWHM apertures.
Results are however compatible with the standard procedure, with a low flux in the $w1$ filter.
Another check was required because the source colour, $b-v \sim 0.7$, is out of the range to which the average count rate to flux ratios (CFR) estimated by \cite{Breeveld} are applicable. Therefore, we explored UVOT calibration issues following the procedure described in \cite{rai10}. We fit the source spectrum with a power law and convolved it with the filter effective areas and appropriate physical quantities to derive source-dependent effective wavelengths (EW) and CFR. However, the results obtained are very similar to those given by Breeveld et al., the largest variations being an increase of the EW by 2 and 3 per cent in the $w1$ and $w2$ band, respectively, and of the CFR by 4 per cent in $w1$. More significant differences were found when comparing the convolved Galactic extinction values with those obtained by simply evaluating them at the filter EW. We obtained a 7 per cent decrease of the extinction in the $w2$ and $m2$ bands, and an 8 per cent increase in the $w1$ band. 
The optical part of the SED remained essentially unchanged after the re-calibration procedure.
The results after the re-calibration procedure are shown in Fig. \ref{fig:SED}.

\subsection{KVA}
\label{KVA}
The KVA telescope is located at the Observatorio del Roque de los Muchachos, La Palma (Canary Islands, Spain), and is operated by the Tuorla Observatory, Finland\footnote{\href{http://users.utu.fi/kani/1m}{http://users.utu.fi/kani/1m}}. The telescope consists of a 0.6-m f/15 Cassegrain devoted to polarimetry, and a 0.35-m f/11 Schmidt-Cassegrain auxiliary telescope for multicolour photometry. The telescope has been successfully operated remotely since the fall of 2003. The KVA is used for optical support observations for MAGIC by making R-band photometric observations, typically one measurement per night per source. \newline
H1722+119 has been regularly monitored by the KVA since 2005. At the beginning of May 2013, after an extended optical high state, the source reached an R-band magnitude of 14.65 (flux of $5.96 \pm 0.09$\,mJy), which constituted a historical maximum for this source at that time\footnote{The highest flux of $7.65 \pm 0.11$\,mJy was measured in 2014 June, and the lowest in 2008 April ($2.18 \pm 0.05$\,mJy).}. 
The high emission state triggered observations by MAGIC, but the MAGIC observations started during the decreasing part of the optical flaring activity. The KVA observed nightly light curve for 2013 is shown in Fig.~\ref{fig:MWL}. The flux varied significantly, the ratio between the highest and lowest flux being 1.6, but we saw no regularity in this variation. \newline
The data were reduced by the Tuorla Observatory Team as described in Nilsson et al. (in preparation). The data were corrected for Galactic extinction using the total absorption $A_{\lambda} = 0.374$\,mag from \citet{schlafly11}.

\subsection{OVRO}
\label{OVRO}
The 40-m radio telescope at the OVRO observes at the 15 GHz band. In late 2007, it started regular monitoring of a sample of blazars in order to support the goals of the {\it Fermi}-LAT telescope \citep{OVRO}. This monitoring programme includes about 1800 known or potential $\gamma$-ray-loud blazars, including all candidate $\gamma$-ray blazar survey \citep[CGRaBS;][]{CGRaBS} sources above declination $-20^\circ$. The sources in this programme are observed in total intensity twice per week with a 4\,mJy (minimum) and 3 per cent (typical) uncertainty on the flux density. Observations are performed with a dual-beam (each 2.5 arcmin full-width half-maximum) Dicke-switched system using cold sky in the off-source beam as the reference. Additionally, the source is switched between beams to reduce atmospheric variations. The absolute flux density scale is calibrated using observations of 3C 286, adopting the flux density (3.44\,Jy) from \citet{Baars1977}. This results in about a 5 per cent absolute scale uncertainty, which is not reflected in the plotted errors in Fig. \ref{fig:MWL}. \newline
The OVRO nightly light curve for 2013 is shown in Fig.~\ref{fig:MWL}. Although there is some indication of a short time variability, we fitted the whole sample with a linear function ($F\mbox{[Jy]} = p_0 + p_1 \times (t - 56300)\mbox{[MJD]}$) to point out the general trend of increasing flux. The fit parameters are $p_0 = (5.2 \pm 0.2)\times 10^{-2}$\,Jy, $p_1 = (1.07 \pm 0.10)\times 10^{-4}$\,Jy/day and $\chi^2 / NDF = 39.68/36$. We also considered the possibility of a constant flux, but that assumption was discarded with $\chi^2 / NDF = 146.9/37$ ($p = 5\times10^{-15}$).

\section{Results}
\label{MWL}

\subsection{Redshift and the intrinsic VHE $\gamma$-ray spectrum}
\label{Redshift}
VHE $\gamma$-rays can be absorbed by the extragalactic background light (EBL), through photon-photon interactions, resulting in $e^+e^-$ pair-production. The flux attenuation is directly dependent on the redshift of the source and energy of $\gamma$-rays. A redshift-estimation method described in \citet{PrandiniMethod} uses this fact. It relies on the assumption that both HE and VHE $\gamma$-rays are created by the same physical processes and in the same region, and that the intrinsic spectrum in the VHE range cannot be harder than the spectrum in the HE range. 
The method uses only the spectral slope of the HE range data. The VHE $\gamma$-ray spectrum was de-absorbed using the EBL model from \citet{Franceschini}. 
Because spectral points obtained by the MAGIC telescopes at energies below 100\,GeV are usually affected by larger systematic uncertainties, they are not used for the fit of the VHE spectrum.
The reconstructed redshift is obtained by applying a simple empirical formula to the value estimated from de-absorption.
Applying this method to the MAGIC and contemporaneous (2013 May; MJD 56413 -- 56442) {\it Fermi}-LAT data we obtained the reconstructed redshift of H1722+119 to be $z = 0.34 \pm 0.15_{\rm stat} \pm 0.05_{\rm meth}$, the error marked as $meth$ being a result of the method as described in \citet{PrandiniMethod}. The UL on the redshift was set to 1.06 for the 95 per cent confidence level. Applying the log likelihood ratio test to set the UL, as described in e.g. \citet{Mazin2007}, we obtained a value of 0.95. When de-absorbed for redshift values greater than 0.95, the spectrum shape becomes parabolic in a $\log(dN/dE)$ vs $\log E$ representation, 
with apparent minimum at $\approx200$\,GeV. 
Our reconstructed redshift value is in agreement with the latest \citet{Landoni2014} and Farina et al. (2013) results. We used our result combined with the LL from Farina et al. $(z > 0.4)$ to de-absorb the VHE $\gamma$-ray flux. For $z = 0.4$, and using the EBL model from \citet{Franceschini}, we found the parameters of the intrinsic (EBL-deabsorbed) VHE spectrum to be $f_0 = (9.6 \pm 2.2) \times 10^{-11}\,\rm ph\,cm^{-2}\,s^{-1}\,TeV^{-1}$, $\rm \Gamma = 2.3 \pm 0.4$, $\chi ^2 / NDF = 3.1/8$.

\subsection{Multifrequency light curve}
\begin{figure*}
  \vbox to220mm{
  \includegraphics[width=18.0cm]{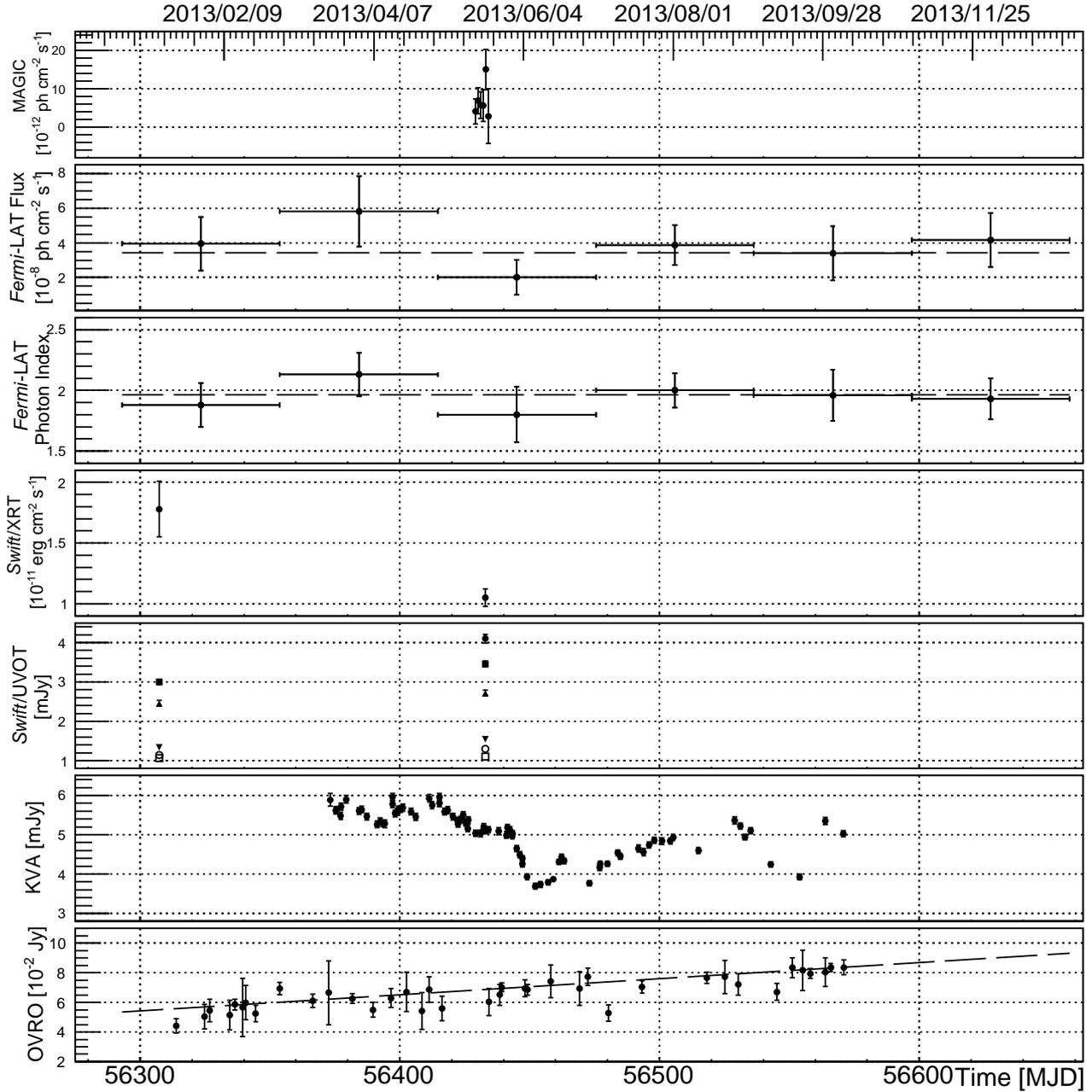}
  \caption{Multifrequency light curve during 2013. MAGIC nightly light curve is the same as in Fig.~\ref{fig:MAGIC_LC}. ULs were omitted for clarity. The HE $\gamma$-ray light curve and the evolution of the spectral index, as measured by {\it Fermi}-LAT, are shown in 2-month time bins, while one night time-scale is used in light curves in other energy bands. Horizontal dashed lines in both panels represent the fit with a constant. The {\it Swift}/UVOT filters are represented with the following markers: $v$ -- full circle, $b$ -- full square, $u$ -- upward triangle, $w1$ -- downward triangle, $m2$ -- empty circle, and $w2$ -- empty square. The dashed line in the OVRO panel represents a linear function fit. All the fit parameters are given in the text. 
The data were collected (from top to bottom) by MAGIC ($E > 150$\,GeV), {\it Fermi}-LAT ($0.1\,{\rm GeV} < E < 100$\,GeV), {\it Swift}/XRT, {\it Swift}/UVOT, KVA (R-band) and OVRO (15\,GHz).}
 \label{fig:MWL}}
\end{figure*}

As already mentioned in Section \ref{KVA}, MAGIC observations were triggered by an extended optical high state, which was the historical R-band maximum at that time. The VHE $\gamma$-ray flux (Fig.~\ref{fig:MAGIC_LC}) is compatible with a constant flux and the previously established UL based on combined data taken over several years. None the less, we cannot reach a firm conclusion on whether the source was flaring in the VHE $\gamma$-rays during MAGIC observations or not. Observations performed during six consecutive nights were not sufficient to study a longer term variability. However, we compared the HE $\gamma$-ray light curve for the entire year to the optical light curve over period 2013 March 22 to October 05, and the radio light curve over period 2013 January 21 to October 05. The {\it Fermi}-LAT data were divided into 2-month time bins with the photon index left free to vary (Fig.~\ref{fig:MWL}), while each point in the KVA and OVRO light curves represents a single measurement. 
Comparison of fluxes for the entire period of collected data revealed that the OVRO data show an obvious trend of increasing flux on a time-scale of one year (dashed line in the bottom panel of Fig.~\ref{fig:MWL}), while the HE $\gamma$-ray flux was consistent with being constant, and the optical flux varied with no apparent regularity. Therefore, we cannot claim any connection between emissions in different energy bands. 

\subsection{SED modelling}

\begin{figure}
\centering
\includegraphics[width=0.5\textwidth]{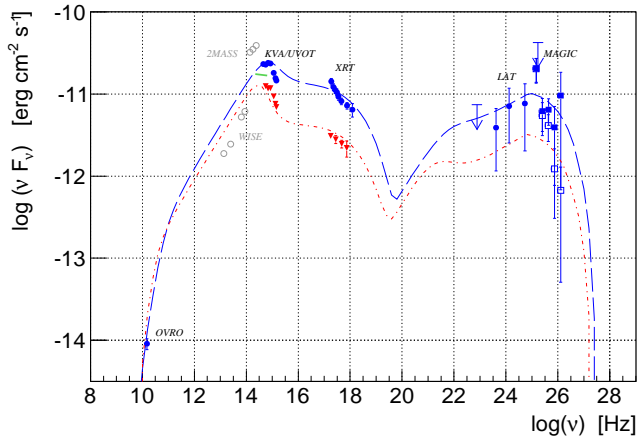}
\caption{The H1722+119 SED. Blue dots represents data contemporaneous to MAGIC observations. The MAGIC measured data are shown by empty blue squares, while full blue squares indicate the de-absorbed points using the EBL model from \citet{Franceschini}. {\em Fermi}-LAT data are shown by dots, while arrows represent {\em Fermi}-LAT ULs. The {\em Swift} data indicated by blue dots, were taken on 2013 May 20 (MJD 56432). The KVA R-band point represented by a blue dot was taken on the same night. The OVRO measurement represented by a blue dot was taken on 2013 May 22 (MJD 56434). Red triangles show the {\em Swift} data from 2008 May 31 (MJD 54617). The green solid line in the $10^{14} - 10^{15}$\,Hz range indicates data from \citet{Landoni2014}, but they were not considered for the fit. Archival data from 2MASS and WISE are shown by empty grey circles, and were not considered for the fit. The blue long-dashed line indicates fits of the helical jet model to the 2013 data, while the fit to the 2008 data is indicated by the red dash-dot line. Both models represent the intrinsic VHE emission. Only blue and red points were considered for modelling.}
\label{fig:SED}
\end{figure}

The SED for 2013 May is shown in Fig.~\ref{fig:SED}. Only data contemporaneous to MAGIC observations have been considered for modelling the SED for 2013. 
On 2013 May 22 (MJD 56434), one observation was performed by the OVRO at 15 GHz. 
The {\it Swift} data collected on 2013 May 20 (MJD 56432) were considered, and the R-band observation by the KVA from the same night was used to obtain a spectral point at $4.56\times10^{14}$\,Hz.
The {\it Fermi}-LAT spectrum was calculated in the period 2013 May 1 -- June 30. 
The MAGIC spectral points were obtained with Schmelling's method as described in \citet{ForwardUnfold}. Based on the arguments discussed in Section \ref{Redshift}, we adopted a value for redshift of 0.4, and EBL model from \citet{Franceschini} to get the intrinsic VHE part of the spectrum. 
We also show the {\it Swift} data collected on 2008 May 31 (MJD 54617), 
in order to compare the source SED during the two different brightness states. 
Archival data from the 2MASS (Two Micron All Sky Survey) All-Sky Catalog of Point Sources \citep{Skrutskie2006} and AllWISE ({\it Wide-field Infrared Survey Explorer}) Data Release \citep{Ochsenbein2000} are included to show the overall behaviour in the IR-optical band, however these were not considered for modelling. We also show the near-IR and optical spectral points from \citet{Landoni2014}, which were also not used for modelling.

The source SED shows surprising behaviour in the frequency range $\sim 3\times10^{14} - 10^{18}$\,Hz. Indeed, the de-absorbed (as described in Section \ref{Swift}) optical--UV spectrum appears curved, with a peak in the $b$ band and a steep slope in the UV. 
H1722+119 is a bona fide BL Lac type of source. None of the observations used in this work, nor information found in the literature revealed anything about the nature of the black hole environment, nor the host galaxy. There is no evidence of emission at any frequency from the accretion disc, dust torus or BLR. In fact, \citet{Landoni2014} detected no intrinsic or intervening spectral lines, and set the ratio of beamed to thermal emission to be $\ge400$.

Therefore we do not expect the presence of thermal emission from an accretion disc and the host galaxy that can justify the shape of the optical--UV spectrum.
Moreover, this shape prevents a monotonic connection with the X-ray spectrum, as would be expected if both the optical--UV and X-ray emissions were produced by a synchrotron process in the same jet region. 
The connection between the UV and X-ray spectrum in both states requires an inflection point that is hard to reproduce with simple, one-zone models. 
This instead can easily be obtained in the framework of a curved jet. Indeed, curved/helical geometries have often been observed in blazar jets \citep[see e.g.][and references therein]{vil99}. 
A helical-jet morphology can arise as the result of perturbations induced by orbital motion in a binary black hole system or precession of the black hole spin axis \bibpunct[, ]{(}{}{;}{a}{}{,}\citep[][and references therein]{vil99}; \bibpunct[, ]{}{)}{;}{a}{}{,}\citep{Rieger2004}\bibpunct[, ]{(}{)}{;}{a}{}{;}. 3-D magnetohydrodinamic (MHD) simulations by \citet{Nakamura2001,Moll2008,Mignone2010} show how kink instabilities lead to a  wiggled, in particular helical, jet structure. MHD equilibrium of helical jets was investigated by \citet{Villata95}.

\begin{figure}
\centering
\includegraphics[width=0.5\textwidth]{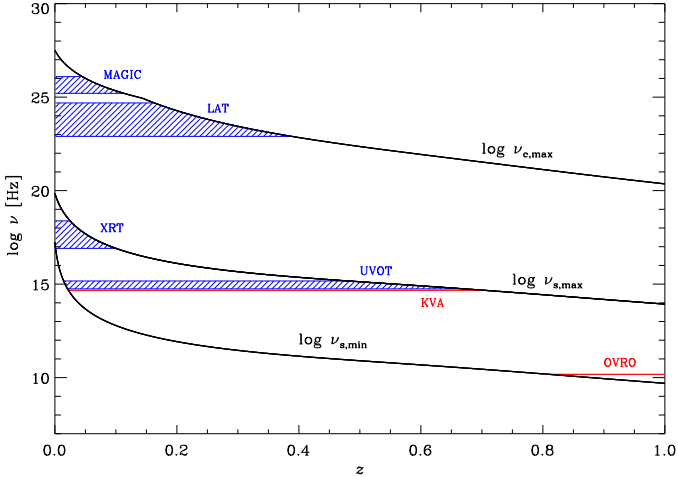}
\caption{The trend of the observed frequencies as a function of the distance along the helical jet axis $z$ (whose unit length can be estimated to be about 0.1 pc, see text). The location of the regions that contribute to the emission observed by the various instruments is highlighted. The plot refers to the high emission state shown in Fig. \ref{fig:SED}.}
\label{fig:jetEmission}
\end{figure}

We adopted the helical jet model originally proposed by \citet{vil99} to explain the observed SED variations of Mkn 501, and later applied also to other objects like 
S4 0954+65 \citep{rai99}, AO 0235+16 \citep{ost04}, BL Lacertae \citep{rai09,rai10} and PG 1553+113 \citep{raiteri2015}. 
The axis of the helical-shaped jet is assumed to lie
along the $z$-axis of a 3-D reference frame. The pitch angle is
$\zeta$ and $\psi$ is the angle defined by the helix axis with the
line of sight. 
The non-dimensional length of the helical path can be expressed in terms
of the $z$ coordinate along the helix axis:
\begin{equation} l(z)=\frac{z}{\cos \zeta}\,,\quad 0\le z \le 1\,,  
\end{equation}
which corresponds to an azimuthal angle 
$\varphi(z)=az$, where the angle $a$ is a constant.
The jet viewing angle varies along the helical path as
\begin{equation}			\label{costeta}
\cos\theta(z)=\cos\psi\cos\zeta+\sin\psi\sin\zeta\cos(\phi- a z)\,,
\end{equation}
where $\phi$ is the azimuthal difference between the line of sight and the
initial direction of the helical path. 
The jet is inhomogeneous: each slice of the jet can radiate, in the plasma rest
reference frame, synchrotron photons from a minimum
frequency $\nu'_{\rm{s,min}}$ to a maximum one $\nu'_{\mathrm{s,max}}$, which follow the laws:
\begin{equation}		 \label{nusyn}
\nu'_{{\rm s},i}(l)=\nu'_{{\rm s},i}(0)
\left(1+\frac{l}{l_i}\right)^{-c_i} \, , \quad c_i>0\,, 
\end{equation}
where $l_i$ are length scales, and $i=\rm min,max$.
The model takes into account IC scattering of the synchrotron photons
by the same relativistic electrons emitting them (i.e. synchrotron self-Compton). 
Consequently, each portion of the jet emitting synchrotron radiation between 
$\nu'_{\rm s,min}(l)$ and $\nu'_{\rm s,max}(l)$ will also produce IC radiation between $\nu'_{\rm c,min}(l)$ and $\nu'_{\rm c,max}(l)$,
with $\nu'_{{\rm c},i}(l)= \frac{4}{3}\gamma_i^{2}(l) \nu'_{{\rm s},i}(l)$.
The electron Lorentz factor ranges from $\gamma_{\rm min}=1$ to $\gamma_{\rm max}(l)$, 
which has a similar dependence as in Eq.\ \ref{nusyn}, 
with power $c_\gamma$ and length scale $l_\gamma$.
As photon energies increase, the classical Thomson scattering cross
section 
gradually shifts into the extreme
Klein-Nishina one, which makes Compton scattering less efficient.
In our approximation
$\nu'_{\rm c,max}(l)$ is averaged with
$\nu'^{\rm KN}_{\rm c,max}(l)=\frac{m_{\rm{e}} c^{2}}{h} \, \gamma_{\rm max}(l)$ 
when $\gamma_{\rm max}(l)\nu'_{\rm s,max}(l) >
\frac{3}{4} \frac{m_{\rm{e}} c^{2}}{h}$. 
We assume a power law dependence of the observed flux density on the frequency and
a cubic dependence on the Doppler beaming factor $\delta$:
$F_\nu(\nu)\propto \delta^3 \nu^{-\alpha_0}$, 
where $\alpha_0$ is the power law index of the local synchrotron spectrum,
$\delta =[\Gamma_{\rm b}(1-\beta\cos\theta)]^{-1}$,
$\beta$ is the bulk velocity of the emitting plasma
in units of the speed of light, $\Gamma_{\rm b}=(1-\beta^2)^{-1/2}$ the corresponding
bulk Lorentz factor, and $\theta$ is the viewing angle of Eq.\ \ref{costeta}.
The variation of the viewing angle along the helical path implies a change of the beaming factor.
As a consequence, the flux at $\nu$ peaks when the part of the jet mostly
contributing to it has minimum $\theta$. 
The emissivity varies along the jet, so that for a jet slice of thickness d$l$:
\begin{equation}
{\rm d}F_{\nu,{\rm s}}(\nu) = K_{\rm s} \, \delta^3(l) \, \nu^{-\alpha_0} \,
\left(1+\frac{l}{l_{\rm s}}\right)^{-c_{\rm s1}} \, \left(\frac{l}{l_{\rm s}}\right)^{c_{\rm s1}/c_{\rm s2}}
{\rm d}l,
\end{equation}
\begin{multline}
{\rm d}F_{\nu,{\rm c}}(\nu) = K_{\rm c} \, \delta^3(l) \, \nu^{-\alpha_0} \,
\left(1+\frac{l}{l_{\rm c}}\right)^{-c_{\rm c1}} \, \left(\frac{l}{l_{\rm c}}\right)^{c_{\rm c1}/c_{\rm c2}}
\\ \times\ln\left[\frac{\nu'_{\rm s,max}(l)}{\nu'_{\rm s,min}(l)}\right]  
{\rm d}l,
\end{multline}
where  $c_{\rm s1},c_{\rm s2},c_{\rm c1},c_{\rm c2}>0$.
The observed flux densities at frequency $\nu$ coming from the whole emitting jet are obtained by integrating over all the jet portions
contributing to that observed frequency (see Fig. \ref{fig:jetEmission}). Therefore, the observed flux density at each frequency will in general differ depending on the emissivity and beaming of each contributing jet portion, resulting in the overall shape of the SED. 
\begin{table}
\caption{Main parameters of the helical model to fit the SEDs of H1722+119. 
The only difference between the high and low state is the angle $\phi$, which changes from $25^{\circ}$ to $31^{\circ}$. }
\label{tab:para}
\centering
\begin{tabular}{l r }
\hline
$\zeta$    & $30^{\circ}$\\
$a$        & $40^{\circ}$\\
$\psi$     & $25^{\circ}$\\
$\phi$     & $25^{\circ}, 31^{\circ}$\\
$\alpha_0$ & 0.5 \\
$\Gamma_{\rm b}$   & 10 \\
$\log\,\nu'_{\rm s,min}(0)$ & 16.7 \\
$\log\,\nu'_{\rm s,max}(0)$ & 19.3 \\
$c_{\rm min,max}$ & 3.3 \\
$\log l_{\rm min}$ & $-2.3$ \\
$\log l_{\rm max}$ & $-1.8$ \\
$\log \gamma_{\rm max}(0)$ & 4.8\\
$c_\gamma$ & 1.5 \\
$\log l_\gamma$ & $-1$ \\
$c_{\rm s1,c1}$ & 3 \\
$c_{\rm s2,c2}$ & 2.3 \\
$\log l_{\rm s,c}$ & $-1$ \\
$\log K_{\rm s}$ & $-19.25$ \\
$\log K_{\rm c}$ & $-24.55$ \\
\hline
\end{tabular}
\end{table}

The helical jet model is intended to be a geometrical, dimensionless, model used to describe blazar variability and should not be interpreted as a physical model. However, absolute dimensions can be derived by comparison with the Very Long Baseline Array (VLBA) images. In our model, the 15\,GHz radio emission comes from the outer 20 per cent of the jet (see Fig. \ref{fig:jetEmission}). This is assumed to be the jet region where the bulk of the observed 15\,GHz radiation comes from.  The results of the MOJAVE programme\footnote{www.physics.purdue.edu/astro/MOJAVE/sourcepages/1722+119.shtml} show that the size of the emitting core is $\lesssim 0.1$\,pc. Therefore the unit length in our model would correspond to a dimension of the order of one tenth of a parsec. 

The helical jet model can produce reasonable fits to the source SEDs (see Fig. \ref{fig:SED}), although the model applied to 2008 data is only constrained by the synchrotron emission, because there are no contemporaneous data in other energy bands. 
The noticeable thing is that both fits were obtained with the same choice of model parameters (see Table \ref{tab:para}), with the exception of the angle $\phi$, which changes from $25^{\circ}$ to $31^{\circ}$ when going from the high to the low state. This underlines how variations of a few degrees in the viewing angle alone may 
account for the observed flux changes. 

\section{Summary and Conclusions}
\label{summary}

MAGIC detected VHE $\gamma$-radiation from BL Lac H1722+119 after observations were triggered by a high flux in R-band measured by the KVA. The MAGIC observations were performed during six consecutive nights and show no flux variability. 
No significant increase of the activity was observed at HE by {\it Fermi}-LAT in 2013 May neither on short time-scales nor compared to the average 2013 flux. An indication of spectral hardening at HE in 2013 May might explain a VHE $\gamma$-ray flux high enough to be detected by the MAGIC telescopes.
Changes in flux were significant in optical and radio data. However, on a time-scale of a year, the radio flux seems to increase, while there was no significant variability of the HE flux. Therefore,  we cannot claim that the two components originate from the same physical region. 
While the radio flux was highest towards the end of the year, the highest optical flux was observed in April, followed by a quite sudden drop, which did not occur in radio data.

The SED shows an interesting feature in the optical--UV band. A break between the optical--UV parts of the spectrum is clearly visible. It also prevents a smooth connection between optical, UV and X-ray bands. We proposed the helical jet model from \citet{vil99} to explain the observed emission.
We found that the difference between the two states can be attributed to a change of a few degrees in the jet orientation.

Further multifrequency observations will be crucial to investigate in detail this new VHE emitting blazar and its emission mechanisms.

\section*{Acknowledgements}
We are grateful to Marco Landoni for help with adopting data from \citet{Landoni2014}. 

We would like to thank the Instituto de Astrof\'{\i}sica de Canarias for the excellent working conditions at the Observatorio del Roque de los Muchachos in La Palma. The financial support of the German BMBF and MPG, the Italian INFN and INAF, the Swiss National Fund SNF, the ERDF under the Spanish MINECO (FPA2012-39502), and the Japanese JSPS and MEXT is gratefully acknowledged. This work was also supported by the Centro de Excelencia Severo Ochoa SEV-2012-0234, CPAN CSD2007-00042, and MultiDark CSD2009-00064 projects of the Spanish Consolider-Ingenio 2010 programme, by grant 268740 of the Academy of Finland, by the Croatian Science Foundation (HrZZ) Project 09/176 and the University of Rijeka Project 13.12.1.3.02, by the DFG Collaborative Research Centers SFB823/C4 and SFB876/C3, and by the Polish MNiSzW grant 745/N-HESS-MAGIC/2010/0.

The \textit{Fermi}-LAT Collaboration acknowledges generous ongoing support
from a number of agencies and institutes that have supported both the
development and the operation of the LAT as well as scientific data
analysis. These include the National Aeronautics and Space Administration
and the Department of Energy in the United States, the Commissariat \`a
l'Energie Atomique and the Centre National de la Recherche Scientifique /
Institut National de Physique Nucl\'eaire et de Physique des Particules in
France, the Agenzia Spaziale Italiana and the Istituto Nazionale di Fisica
Nucleare in Italy, the Ministry of Education, Culture, Sports, Science and
Technology (MEXT), High Energy Accelerator Research Organization (KEK) and
Japan Aerospace Exploration Agency (JAXA) in Japan, and the
K.~A.~Wallenberg Foundation, the Swedish Research Council and the
Swedish National Space Board in Sweden.

Additional support for science analysis during the operations phase from
the following agencies is also gratefully acknowledged: the Istituto
Nazionale di Astrofisica in Italy and the Centre National d'\'Etudes
Spatiales in France.

The OVRO 40-m monitoring programme is supported in part by NASA grants
NNX08AW31G and NNX11A043G, and NSF grants AST-0808050 and AST-1109911.

Antonio Stamerra acknowledges financial support in the frame of the INAF Senior Scientist programme of ASDC and by the Italian Ministry for Research and Scuola Normale Superiore.

Elisa Prandini gratefully acknowledges the financial support of the Marie Heim-Vogtlin grant of the Swiss National Science Foundation.




\Urlmuskip=0mu plus 1mu\relax
\bibliographystyle{mnras}
\bibliography{MAGIC_H1722}    

\begin{thebibliography}{}
\makeatletter
\relax
\def\mn@urlcharsother{\let\do\@makeother \do\$\do\&\do\#\do\^\do\_\do\%\do\~}
\def\mn@doi{\begingroup\mn@urlcharsother \@ifnextchar [ {\mn@doi@}
  {\mn@doi@[]}}
\def\mn@doi@[#1]#2{\def\@tempa{#1}\ifx\@tempa\@empty \href
  {http://dx.doi.org/#2} {doi:#2}\else \href {http://dx.doi.org/#2} {#1}\fi
  \endgroup}
\def\mn@eprint#1#2{\mn@eprint@#1:#2::\@nil}
\def\mn@eprint@arXiv#1{\href {http://arxiv.org/abs/#1} {{\tt arXiv:#1}}}
\def\mn@eprint@dblp#1{\href {http://dblp.uni-trier.de/rec/bibtex/#1.xml}
  {dblp:#1}}
\def\mn@eprint@#1:#2:#3:#4\@nil{\def\@tempa {#1}\def\@tempb {#2}\def\@tempc
  {#3}\ifx \@tempc \@empty \let \@tempc \@tempb \let \@tempb \@tempa \fi \ifx
  \@tempb \@empty \def\@tempb {arXiv}\fi \@ifundefined
  {mn@eprint@\@tempb}{\@tempb:\@tempc}{\expandafter \expandafter \csname
  mn@eprint@\@tempb\endcsname \expandafter{\@tempc}}}

\bibitem[\protect\citeauthoryear{{Acero} et~al.,}{{Acero}
  et~al.}{2015}]{Acero2015}
{Acero} F.,  et~al., 2015, \mn@doi [\apjs] {10.1088/0067-0049/218/2/23}, \href
  {http://adsabs.harvard.edu/abs/2015ApJS..218...23A} {218, 23}

\bibitem[\protect\citeauthoryear{{Ackermann} et~al.,}{{Ackermann}
  et~al.}{2012}]{ackermann12}
{Ackermann} M.,  et~al., 2012, \mn@doi [\apjs] {10.1088/0067-0049/203/1/4},
  \href {http://adsabs.harvard.edu/abs/2012ApJS..203....4A} {203, 4}

\bibitem[\protect\citeauthoryear{{Albert} et~al.,}{{Albert}
  et~al.}{2006}]{Albert2006_Mrk180}
{Albert} J.,  et~al., 2006, \mn@doi [\apjl] {10.1086/508020}, \href
  {http://adsabs.harvard.edu/abs/2006ApJ...648L.105A} {648, L105}

\bibitem[\protect\citeauthoryear{Albert et~al.,}{Albert
  et~al.}{2007a}]{ForwardUnfold}
Albert J.,  et~al., 2007a, \mn@doi [Nucl. Instrum. Meth. Phys. Res. A]
  {10.1016/j.nima.2007.09.048}, 583, 494

\bibitem[\protect\citeauthoryear{{Albert} et~al.,}{{Albert}
  et~al.}{2007b}]{Albert2007_1ES1011}
{Albert} J.,  et~al., 2007b, \mn@doi [\apjl] {10.1086/521982}, \href
  {http://adsabs.harvard.edu/abs/2007ApJ...667L..21A} {667, L21}

\bibitem[\protect\citeauthoryear{Albert et~al.,}{Albert
  et~al.}{2008a}]{Perform_Split}
Albert J.,  et~al., 2008a, \mn@doi [Nucl. Instrum. Meth. Phys. Res. A]
  {10.1016/j.nima.2008.06.043}, 594, 407

\bibitem[\protect\citeauthoryear{{Albert} et~al.,}{{Albert}
  et~al.}{2008b}]{Albert2008_HBL}
{Albert} J.,  et~al., 2008b, \mn@doi [\apj] {10.1086/587499}, \href
  {http://adsabs.harvard.edu/abs/2008ApJ...681..944A} {681, 944}

\bibitem[\protect\citeauthoryear{{Aleksi{\'c}} et~al.,}{{Aleksi{\'c}}
  et~al.}{2011}]{MAGICStacked}
{Aleksi{\'c}} J.,  et~al., 2011, \mn@doi [\apj] {10.1088/0004-637X/729/2/115},
  \href {http://adsabs.harvard.edu/abs/2011ApJ...729..115A} {729, 115}

\bibitem[\protect\citeauthoryear{Aleksi\'{c} et~al.,}{Aleksi\'{c}
  et~al.}{2012a}]{Perform_Stereo}
Aleksi\'{c} J.,  et~al., 2012a, \mn@doi [Astroparticle Physics]
  {10.1016/j.astropartphys.2011.11.007}, 35, 435

\bibitem[\protect\citeauthoryear{Aleksi\'{c} et~al.,}{Aleksi\'{c}
  et~al.}{2012b}]{Aleksic2012_B32247}
Aleksi\'{c} J.,  et~al., 2012b, \mn@doi [\aap] {10.1051/0004-6361/201117967},
  539, A118

\bibitem[\protect\citeauthoryear{Aleksi\'{c} et~al.,}{Aleksi\'{c}
  et~al.}{2012c}]{Aleksic2012_1ES1215}
Aleksi\'{c} J.,  et~al., 2012c, \mn@doi [\aap] {10.1051/0004-6361/201219133},
  544, A142

\bibitem[\protect\citeauthoryear{Aleksi\'{c} et~al.,}{Aleksi\'{c}
  et~al.}{2014}]{Aleksic2014_1ES1727}
Aleksi\'{c} J.,  et~al., 2014, \mn@doi [\aap] {10.1051/0004-6361/201321360},
  563, A90

\bibitem[\protect\citeauthoryear{Aleksi\'{c} et~al.,}{Aleksi\'{c}
  et~al.}{2015}]{Aleksic2015_1ES0806}
Aleksi\'{c} J.,  et~al., 2015, \mn@doi [\mnras] {10.1093/mnras/stv895}, 451,
  739

\bibitem[\protect\citeauthoryear{Aleksi\'{c} et~al.,}{Aleksi\'{c}
  et~al.}{2016a}]{MAGICperform_1}
Aleksi\'{c} J.,  et~al., 2016a, \mn@doi [Astroparticle Physics]
  {10.1016/j.astropartphys.2015.04.004}, 72, 61

\bibitem[\protect\citeauthoryear{Aleksi\'{c} et~al.,}{Aleksi\'{c}
  et~al.}{2016b}]{MAGICperform_2}
Aleksi\'{c} J.,  et~al., 2016b, \mn@doi [Astroparticle Physics]
  {10.1016/j.astropartphys.2015.02.005}, 72, 76

\bibitem[\protect\citeauthoryear{{Anderhub} et~al.,}{{Anderhub}
  et~al.}{2009}]{Anderhub2009_S50716}
{Anderhub} H.,  et~al., 2009, \mn@doi [\apjl] {10.1088/0004-637X/704/2/L129},
  \href {http://adsabs.harvard.edu/abs/2009ApJ...704L.129A} {704, L129}

\bibitem[\protect\citeauthoryear{Atwood et~al.,}{Atwood
  et~al.}{2009}]{atwood09}
Atwood W.~B.,  et~al., 2009, \mn@doi [\apj] {10.1088/0004-637X/697/2/1071},
  \href {http://adsabs.harvard.edu/abs/2009ApJ...697.1071A} {697, 1071}

\bibitem[\protect\citeauthoryear{{Baars}, {Genzel}, {Pauliny-Toth}  \&
  {Witzel}}{{Baars} et~al.}{1977}]{Baars1977}
{Baars} J.~W.~M.,  {Genzel} R.,  {Pauliny-Toth} I.~I.~K.,   {Witzel} A.,  1977,
  \aap, \href {http://adsabs.harvard.edu/abs/1977A%26A....61...99B} {61, 99}

\bibitem[\protect\citeauthoryear{Barthelmy et~al.,}{Barthelmy
  et~al.}{2005}]{barthelmy05}
Barthelmy S.,  et~al., 2005, \mn@doi [\ssr] {10.1007/s11214-005-5096-3}, 120,
  143

\bibitem[\protect\citeauthoryear{{Baumgartner}, {Tueller}, {Markwardt},
  {Skinner}, {Barthelmy}, {Mushotzky}, {Evans}  \& {Gehrels}}{{Baumgartner}
  et~al.}{2013}]{baumgartner13}
{Baumgartner} W.~H.,  {Tueller} J.,  {Markwardt} C.~B.,  {Skinner} G.~K.,
  {Barthelmy} S.,  {Mushotzky} R.~F.,  {Evans} P.~A.,   {Gehrels} N.,  2013,
  \mn@doi [\apjs] {10.1088/0067-0049/207/2/19}, \href
  {http://adsabs.harvard.edu/abs/2013ApJS..207...19B} {207, 19}

\bibitem[\protect\citeauthoryear{{Bessell} \& {Brett}}{{Bessell} \&
  {Brett}}{1988}]{bessel88}
{Bessell} M.~S.,  {Brett} J.~M.,  1988, \mn@doi [\pasp] {10.1086/132281}, \href
  {http://adsabs.harvard.edu/abs/1988PASP..100.1134B} {100, 1134}

\bibitem[\protect\citeauthoryear{{B{\"o}ttcher}, {Reimer}, {Sweeney}  \&
  {Prakash}}{{B{\"o}ttcher} et~al.}{2013}]{Boettcher2013}
{B{\"o}ttcher} M.,  {Reimer} A.,  {Sweeney} K.,   {Prakash} A.,  2013, \mn@doi
  [\apj] {10.1088/0004-637X/768/1/54}, \href
  {http://adsabs.harvard.edu/abs/2013ApJ...768...54B} {768, 54}

\bibitem[\protect\citeauthoryear{Breeveld, Landsman, Holland, Roming, Kuin  \&
  Page}{Breeveld et~al.}{2011}]{Breeveld}
Breeveld A.~A.,  Landsman W.,  Holland S.~T.,  Roming P.,  Kuin N. P.~M.,
  Page M.~J.,  2011, \mn@doi [AIP Conference Proceedings] {10.1063/1.3621807},
  1358, 373

\bibitem[\protect\citeauthoryear{Brissenden, Tuohy, Remillard, Schwartz  \&
  Hertz}{Brissenden et~al.}{1990}]{Brissenden}
Brissenden R. J.~V.,  Tuohy I.~R.,  Remillard R.~A.,  Schwartz D.~A.,   Hertz
  P.~L.,  1990, \apjs, \href
  {http://adsabs.harvard.edu/abs/1990ApJ...350..578B} {350, 578}

\bibitem[\protect\citeauthoryear{Burrows et~al.,}{Burrows
  et~al.}{2005}]{burrows05}
Burrows D.~N.,  et~al., 2005, \mn@doi [\ssr] {10.1007/s11214-005-5097-2}, 120,
  165

\bibitem[\protect\citeauthoryear{{Cortina}}{{Cortina}}{2013}]{MAGICAtel}
{Cortina} J.,  2013, The Astronomer's Telegram, \href
  {http://adsabs.harvard.edu/abs/2013ATel.5080....1C} {5080, 1}

\bibitem[\protect\citeauthoryear{{Cortina} et~al.,}{{Cortina}
  et~al.}{2005}]{Perform_First}
{Cortina} J.,  et~al., 2005, {Proc. 29th Int. Cosm. Ray Conf., Vol. 5.}, \href
  {http://adsabs.harvard.edu/abs/2005ICRC....5..359C} {5, 359}

\bibitem[\protect\citeauthoryear{Costamante \& Ghisellini}{Costamante \&
  Ghisellini}{2002}]{Costamante_Ghisellini}
Costamante L.,  Ghisellini G.,  2002, \aap, \href
  {http://adsabs.harvard.edu/abs/2002A%26A...384...56C} {384, 56}

\bibitem[\protect\citeauthoryear{Donato, Ghisellini, Tagliaferri  \&
  Fossati}{Donato et~al.}{2001}]{Donato}
Donato D.,  Ghisellini G.,  Tagliaferri G.,   Fossati G.,  2001, \aap, \href
  {http://adsabs.harvard.edu/abs/2001A%26A...375..739D} {375, 739}

\bibitem[\protect\citeauthoryear{Dunlop, McLure, Kukula, Baum, O'Dea  \&
  Hughes}{Dunlop et~al.}{2003}]{Dunlop03}
Dunlop J.~S.,  McLure R.~J.,  Kukula M.~J.,  Baum S.~A.,  O'Dea C.~P.,   Hughes
  D.~H.,  2003, \mn@doi [\mnras] {10.1046/j.1365-8711.2003.06333.x}, 340, 1095

\bibitem[\protect\citeauthoryear{Falomo, Bersanelli, Bouchet  \& Tanzi}{Falomo
  et~al.}{1993}]{Falomo1}
Falomo R.,  Bersanelli M.,  Bouchet P.,   Tanzi E.~G.,  1993, \aj, \href
  {http://adsabs.harvard.edu/abs/1993AJ....106...11F} {106, 11}

\bibitem[\protect\citeauthoryear{Falomo, Scarpa  \& Bersanelli}{Falomo
  et~al.}{1994}]{Falomo2}
Falomo R.,  Scarpa R.,   Bersanelli M.,  1994, \apjs, \href
  {http://adsabs.harvard.edu/abs/1994ApJS...93..125F} {93, 125}

\bibitem[\protect\citeauthoryear{{Fitzpatrick}}{{Fitzpatrick}}{1999}]{fitzpatrick99}
{Fitzpatrick} E.~L.,  1999, \mn@doi [\pasp] {10.1086/316293}, \href
  {http://adsabs.harvard.edu/abs/1999PASP..111...63F} {111, 63}

\bibitem[\protect\citeauthoryear{Fomin, Stepanian, Lamb, Lewis, Punch  \&
  Weekes}{Fomin et~al.}{1994}]{wobble}
Fomin V.~P.,  Stepanian A.~A.,  Lamb R.~C.,  Lewis D.~A.,  Punch M.,   Weekes
  T.~C.,  1994, Astropart. Phys., \href
  {http://www.sciencedirect.com/science/article/pii/0927650594900361} {2, 137}

\bibitem[\protect\citeauthoryear{Forman, Jones, Cominsky, Julien, Murray,
  Peters, Tananbaum  \& Giacconi}{Forman et~al.}{1978}]{FormanUhuru}
Forman W.,  Jones C.,  Cominsky L.,  Julien P.,  Murray S.,  Peters G.,
  Tananbaum H.,   Giacconi R.,  1978, \apjs, \href
  {http://adsabs.harvard.edu/abs/1978ApJS...38..357F} {38, 357}

\bibitem[\protect\citeauthoryear{Franceschini, Rodighiero  \&
  Vaccari}{Franceschini et~al.}{2008}]{Franceschini}
Franceschini A.,  Rodighiero G.,   Vaccari M.,  2008, \aap, \href
  {http://www.aanda.org/index.php?option=com_article&access=bibcode&Itemid=129&bibcode=2008A%2526A...487..837FFUL}
  {487, 837}

\bibitem[\protect\citeauthoryear{{Gehrels} et~al.,}{{Gehrels}
  et~al.}{2004}]{gehrels04}
{Gehrels} N.,  et~al., 2004, \mn@doi [\apj] {10.1086/422091}, \href
  {http://adsabs.harvard.edu/abs/2004ApJ...611.1005G} {611, 1005}

\bibitem[\protect\citeauthoryear{{Ghisellini}, {Tavecchio}, {Foschini},
  {Ghirlanda}, {Maraschi}  \& {Celotti}}{{Ghisellini}
  et~al.}{2010}]{Ghisellini2010}
{Ghisellini} G.,  {Tavecchio} F.,  {Foschini} L.,  {Ghirlanda} G.,  {Maraschi}
  L.,   {Celotti} A.,  2010, \mn@doi [\mnras]
  {10.1111/j.1365-2966.2009.15898.x}, \href
  {http://adsabs.harvard.edu/abs/2010MNRAS.402..497G} {402, 497}

\bibitem[\protect\citeauthoryear{{Goebel}, {Bartko}, {Carmona}, {Galante},
  {Jogler}, {Mirzoyan}, {Coarasa}  \& {Teshima}}{{Goebel}
  et~al.}{2008}]{Perform_2GHz}
{Goebel} F.,  {Bartko} H.,  {Carmona} E.,  {Galante} N.,  {Jogler} T.,
  {Mirzoyan} R.,  {Coarasa} J.~A.,   {Teshima} M.,  2008, International Cosmic
  Ray Conference, \href {http://adsabs.harvard.edu/abs/2008ICRC....3.1481G} {3,
  1481}

\bibitem[\protect\citeauthoryear{Griffiths, Wilson, Ward, Tapia  \&
  Ulvestad}{Griffiths et~al.}{1989}]{Griffiths}
Griffiths R.~E.,  Wilson A.~S.,  Ward M.~J.,  Tapia S.,   Ulvestad J.~S.,
  1989, \mnras, \href {http://adsabs.harvard.edu/abs/1989MNRAS.240...33G} {240,
  33}

\bibitem[\protect\citeauthoryear{{Healey} et~al.,}{{Healey}
  et~al.}{2008}]{CGRaBS}
{Healey} S.~E.,  et~al., 2008, \mn@doi [\apjs] {10.1086/523302}, \href
  {http://adsabs.harvard.edu/abs/2008ApJS..175...97H} {175, 97}

\bibitem[\protect\citeauthoryear{Kalberla, Burton, Hartmann, Arnal, Bajaja,
  Morras  \& P\"{o}ppel}{Kalberla et~al.}{2005}]{kalberla05}
Kalberla P. M.~W.,  Burton W.~B.,  Hartmann D.,  Arnal E.~M.,  Bajaja E.,
  Morras R.,   P\"{o}ppel W. G.~L.,  2005, \mn@doi [\aap]
  {10.1051/0004-6361:20041864}, 440, 775

\bibitem[\protect\citeauthoryear{{Komatsu} et~al.,}{{Komatsu}
  et~al.}{2011}]{Komatsu2011}
{Komatsu} E.,  et~al., 2011, \mn@doi [\apjs] {10.1088/0067-0049/192/2/18},
  \href {http://adsabs.harvard.edu/abs/2011ApJS..192...18K} {192, 18}

\bibitem[\protect\citeauthoryear{Landoni, Falomo, Treves  \&
  Sbarufatti}{Landoni et~al.}{2014}]{Landoni2014}
Landoni M.,  Falomo R.,  Treves A.,   Sbarufatti B.,  2014, \mn@doi [\aap]
  {10.1051/0004-6361/201424232}, 570, A126

\bibitem[\protect\citeauthoryear{Li \& Ma}{Li \& Ma}{1983}]{LiMa}
Li T.-P.,  Ma Y.-Q.,  1983, \apjs, \href {http://arxiv.org/abs/0907.0973} {272,
  317}

\bibitem[\protect\citeauthoryear{Linford, Taylor, Romani, Helmboldt, Readhead,
  Reeves  \& Richards}{Linford et~al.}{2012}]{Linford1}
Linford J.~D.,  Taylor G.~B.,  Romani R.~W.,  Helmboldt J.~F.,  Readhead A.
  C.~S.,  Reeves R.,   Richards J.~L.,  2012, \apjs, \href
  {http://adsabs.harvard.edu/abs/2012ApJ...744..177L} {744, 177}

\bibitem[\protect\citeauthoryear{Lister et~al.,}{Lister
  et~al.}{2011}]{MOJAVEProgram}
Lister M.~L.,  et~al., 2011, \apjs, \href
  {http://adsabs.harvard.edu/abs/2011ApJ...742...27L} {742, 27}

\bibitem[\protect\citeauthoryear{Lombardi, Berger, Colin, Ortega, Klepser  \&
  {for the MAGIC Collaboration}}{Lombardi et~al.}{2011}]{Lombardi}
Lombardi S.,  Berger K.,  Colin P.,  Ortega A.~D.,  Klepser S.,   {for the
  MAGIC Collaboration} 2011, \mn@doi [Proc. Int. Cosm. Ray Conf.]
  {10.7529/ICRC2011/V03/1150}, \href
  {http://adsabs.harvard.edu/abs/2011ICRC....3..266L} {3, 266}

\bibitem[\protect\citeauthoryear{Mao}{Mao}{2011}]{2MASS}
Mao L.~S.,  2011, \na, \href
  {http://adsabs.harvard.edu/abs/2011NewA...16..503M} {16, 503}

\bibitem[\protect\citeauthoryear{{Mattox} et~al.,}{{Mattox}
  et~al.}{1996}]{mattox96}
{Mattox} J.~R.,  et~al., 1996, \mn@doi [\apj] {10.1086/177068}, \href
  {http://adsabs.harvard.edu/abs/1996ApJ...461..396M} {461, 396}

\bibitem[\protect\citeauthoryear{{Mazin} \& {Goebel}}{{Mazin} \&
  {Goebel}}{2007}]{Mazin2007}
{Mazin} D.,  {Goebel} F.,  2007, \mn@doi [\apjl] {10.1086/511751}, 655, L13

\bibitem[\protect\citeauthoryear{{Mignone}, {Rossi}, {Bodo}, {Ferrari}  \&
  {Massaglia}}{{Mignone} et~al.}{2010}]{Mignone2010}
{Mignone} A.,  {Rossi} P.,  {Bodo} G.,  {Ferrari} A.,   {Massaglia} S.,  2010,
  \mn@doi [\mnras] {10.1111/j.1365-2966.2009.15642.x}, \href
  {http://adsabs.harvard.edu/abs/2010MNRAS.402....7M} {402, 7}

\bibitem[\protect\citeauthoryear{{Moll}, {Spruit}  \& {Obergaulinger}}{{Moll}
  et~al.}{2008}]{Moll2008}
{Moll} R.,  {Spruit} H.~C.,   {Obergaulinger} M.,  2008, \mn@doi [\aap]
  {10.1051/0004-6361:200810523}, \href
  {http://adsabs.harvard.edu/abs/2008A%26A...492..621M} {492, 621}

\bibitem[\protect\citeauthoryear{{Nakamura}, {Uchida}  \& {Hirose}}{{Nakamura}
  et~al.}{2001}]{Nakamura2001}
{Nakamura} M.,  {Uchida} Y.,   {Hirose} S.,  2001, \mn@doi [\na]
  {10.1016/S1384-1076(01)00041-0}, \href
  {http://adsabs.harvard.edu/abs/2001NewA....6...61N} {6, 61}

\bibitem[\protect\citeauthoryear{{Nieppola}, {Tornikoski}  \&
  {Valtaoja}}{{Nieppola} et~al.}{2006}]{NieppolaCatalog}
{Nieppola} E.,  {Tornikoski} M.,   {Valtaoja} E.,  2006, \mn@doi [\aap]
  {10.1051/0004-6361:20053316}, \href
  {http://adsabs.harvard.edu/abs/2006A%26A...445..441N} {445, 441}

\bibitem[\protect\citeauthoryear{{Ochsenbein}, {Bauer}  \&
  {Marcout}}{{Ochsenbein} et~al.}{2000}]{Ochsenbein2000}
{Ochsenbein} F.,  {Bauer} P.,   {Marcout} J.,  2000, \mn@doi [\aaps]
  {10.1051/aas:2000169}, \href
  {http://adsabs.harvard.edu/abs/2000A%26AS..143...23O} {143, 23}

\bibitem[\protect\citeauthoryear{{Ostorero}, {Villata}  \&
  {Raiteri}}{{Ostorero} et~al.}{2004}]{ost04}
{Ostorero} L.,  {Villata} M.,   {Raiteri} C.~M.,  2004, \mn@doi [\aap]
  {10.1051/0004-6361:20035813}, \href
  {http://adsabs.harvard.edu/cgi-bin/nph-bib_query?bibcode=2004A%26A...419..913O&db_key=AST}
  {419, 913}

\bibitem[\protect\citeauthoryear{Prandini, Bonnoli, Maraschi, Mariotti  \&
  Tavecchio}{Prandini et~al.}{2010}]{PrandiniMethod}
Prandini E.,  Bonnoli G.,  Maraschi L.,  Mariotti M.,   Tavecchio F.,  2010,
  \mn@doi [\mnras] {10.1111/j.1745-3933.2010.00862.x}, 405, L76

\bibitem[\protect\citeauthoryear{{Raiteri} et~al.,}{{Raiteri}
  et~al.}{1999}]{rai99}
{Raiteri} C.~M.,  et~al., 1999, \aap, \href
  {http://adsabs.harvard.edu/cgi-bin/nph-bib_query?bibcode=1999A%26A...352...19R&db_key=AST}
  {352, 19}

\bibitem[\protect\citeauthoryear{{Raiteri} et~al.,}{{Raiteri}
  et~al.}{2009}]{rai09}
{Raiteri} C.~M.,  et~al., 2009, \mn@doi [\aap] {10.1051/0004-6361/200912953},
  \href {http://adsabs.harvard.edu/abs/2009A%26A...507..769R} {507, 769}

\bibitem[\protect\citeauthoryear{{Raiteri} et~al.,}{{Raiteri}
  et~al.}{2010}]{rai10}
{Raiteri} C.~M.,  et~al., 2010, \mn@doi [\aap] {10.1051/0004-6361/201015191},
  \href {http://adsabs.harvard.edu/abs/2010A%26A...524A..43R} {524, A43}

\bibitem[\protect\citeauthoryear{{Raiteri} et~al.,}{{Raiteri}
  et~al.}{2015}]{raiteri2015}
{Raiteri} C.~M.,  et~al., 2015, \mn@doi [\mnras] {10.1093/mnras/stv1884}, \href
  {http://adsabs.harvard.edu/abs/2015MNRAS.454..353R} {454, 353}

\bibitem[\protect\citeauthoryear{Richards et~al.,}{Richards
  et~al.}{2011}]{OVRO}
Richards J.~L.,  et~al., 2011, \apjs, \href
  {http://adsabs.harvard.edu/abs/2011ApJS..194...29R} {194, 29}

\bibitem[\protect\citeauthoryear{{Rieger}}{{Rieger}}{2004}]{Rieger2004}
{Rieger} F.~M.,  2004, \mn@doi [\apjl] {10.1086/426018}, \href
  {http://adsabs.harvard.edu/abs/2004ApJ...615L...5R} {615, L5}

\bibitem[\protect\citeauthoryear{Roming et~al.,}{Roming
  et~al.}{2005}]{roming05}
Roming P.,  et~al., 2005, \mn@doi [\ssr] {10.1007/s11214-005-5095-4}, 120, 95

\bibitem[\protect\citeauthoryear{{Sbarufatti}, {Treves}  \&
  {Falomo}}{{Sbarufatti} et~al.}{2005}]{Sbarufatti2005}
{Sbarufatti} B.,  {Treves} A.,   {Falomo} R.,  2005, \mn@doi [\apj]
  {10.1086/497022}, \href {http://adsabs.harvard.edu/abs/2005ApJ...635..173S}
  {635, 173}

\bibitem[\protect\citeauthoryear{Sbarufatti, Treves, Falomo, Heidt, Kotilainen
  \& Scarpa}{Sbarufatti et~al.}{2006}]{Sbarufatti}
Sbarufatti B.,  Treves A.,  Falomo R.,  Heidt J.,  Kotilainen J.,   Scarpa R.,
  2006, \aj, \href {http://adsabs.harvard.edu/abs/2006AJ....132....1S} {132, 1}

\bibitem[\protect\citeauthoryear{{Schlafly} \& {Finkbeiner}}{{Schlafly} \&
  {Finkbeiner}}{2011}]{schlafly11}
{Schlafly} E.~F.,  {Finkbeiner} D.~P.,  2011, \mn@doi [\apj]
  {10.1088/0004-637X/737/2/103}, \href
  {http://adsabs.harvard.edu/abs/2011ApJ...737..103S} {737, 103}

\bibitem[\protect\citeauthoryear{Sitarek, Gaug, Mazin, Paoletti  \&
  Tescaro}{Sitarek et~al.}{2013}]{Perform_DRS4}
Sitarek J.,  Gaug M.,  Mazin D.,  Paoletti R.,   Tescaro D.,  2013, \mn@doi
  [Nucl. Instrum. Meth. Phys. Res. A] {10.1016/j.nima.2013.05.014}, 723, 109

\bibitem[\protect\citeauthoryear{{Skrutskie} et~al.,}{{Skrutskie}
  et~al.}{2006}]{Skrutskie2006}
{Skrutskie} M.~F.,  et~al., 2006, \mn@doi [\aj] {10.1086/498708}, \href
  {http://adsabs.harvard.edu/abs/2006AJ....131.1163S} {131, 1163}

\bibitem[\protect\citeauthoryear{V\'{e}ron-Cetty \& V\'{e}ron}{V\'{e}ron-Cetty
  \& V\'{e}ron}{1993}]{VeronCetty}
V\'{e}ron-Cetty M.-P.,  V\'{e}ron P.,  1993, \aaps, \href
  {http://adsabs.harvard.edu/abs/1993A%26AS..100..521V} {100, 521}

\bibitem[\protect\citeauthoryear{{Villata} \& {Ferrari}}{{Villata} \&
  {Ferrari}}{1995}]{Villata95}
{Villata} M.,  {Ferrari} A.,  1995, \aap, \href
  {http://adsabs.harvard.edu/abs/1995A%26A...293..626V} {293, 626}

\bibitem[\protect\citeauthoryear{{Villata} \& {Raiteri}}{{Villata} \&
  {Raiteri}}{1999}]{vil99}
{Villata} M.,  {Raiteri} C.~M.,  1999, \aap, \href
  {http://adsabs.harvard.edu/cgi-bin/nph-bib_query?bibcode=1999A%26A...347...30V&db_key=AST}
  {347, 30}

\bibitem[\protect\citeauthoryear{{Wilms}, {Allen}  \& {McCray}}{{Wilms}
  et~al.}{2000}]{wilms00}
{Wilms} J.,  {Allen} A.,   {McCray} R.,  2000, \mn@doi [\apj] {10.1086/317016},
  \href {http://adsabs.harvard.edu/abs/2000ApJ...542..914W} {542, 914}

\bibitem[\protect\citeauthoryear{Wood et~al.,}{Wood et~al.}{1984}]{HEAOWood}
Wood K.~S.,  et~al., 1984, \apjs, \href
  {http://adsabs.harvard.edu/abs/1984ApJS...56..507W} {56, 507}

\bibitem[\protect\citeauthoryear{Zanin et~al.,}{Zanin et~al.}{2013}]{MARS}
Zanin R.,  et~al., 2013, Proceedings of $33^{rd}$ ICRC, p.~0773

\makeatother
\end{thebibliography}


\vspace*{0.5cm}
\noindent
$^{1}$ {ETH Zurich, CH-8093 Zurich, Switzerland} \\
$^{2}$ {Universit\`a di Udine, and INFN Trieste, I-33100 Udine, Italy} \\
$^{3}$ {INAF National Institute for Astrophysics, I-00136 Rome, Italy} \\
$^{4}$ {Universit\`a  di Siena, and INFN Pisa, I-53100 Siena, Italy} \\
$^{5}$ {Croatian MAGIC Consortium, Rudjer Boskovic Institute, University of Rijeka, University of Split and University of Zagreb, Croatia} \\
$^{6}$ {Saha Institute of Nuclear Physics, 1/AF Bidhannagar, Salt Lake, Sector-1, Kolkata 700064, India} \\
$^{7}$ {Max-Planck-Institut f\"ur Physik, D-80805 M\"unchen, Germany} \\
$^{8}$ {Universidad Complutense, E-28040 Madrid, Spain} \\
$^{9}$ {Inst. de Astrof\'isica de Canarias, E-38200 La Laguna, Tenerife, Spain; Universidad de La Laguna, Dpto. Astrof\'isica, E-38206 La Laguna, Tenerife, Spain} \\
$^{10}$ {University of \L\'od\'z, PL-90236 Lodz, Poland} \\
$^{11}$ {Deutsches Elektronen-Synchrotron (DESY), D-15738 Zeuthen, Germany} \\
$^{12}$ {Institut de Fisica d'Altes Energies (IFAE), The Barcelona Institute of Science and Technology, Campus UAB, 08193 Bellaterra (Barcelona), Spain} \\
$^{13}$ {Universit\"at W\"urzburg, D-97074 W\"urzburg, Germany} \\
$^{14}$ {Universit\`a di Padova and INFN, I-35131 Padova, Italy} \\
$^{15}$ {Institute for Space Sciences (CSIC/IEEC), E-08193 Barcelona, Spain} \\
$^{16}$ {Technische Universit\"at Dortmund, D-44221 Dortmund, Germany} \\
$^{17}$ {Finnish MAGIC Consortium, Tuorla Observatory, University of Turku and Astronomy Division, University of Oulu, Finland} \\
$^{18}$ {Unitat de F\'isica de les Radiacions, Departament de F\'isica, and CERES-IEEC, Universitat Aut\`onoma de Barcelona, E-08193 Bellaterra, Spain} \\
$^{19}$ {Universitat de Barcelona, ICC, IEEC-UB, E-08028 Barcelona, Spain} \\
$^{20}$ {Japanese MAGIC Consortium, ICRR, The University of Tokyo, Department of Physics and Hakubi Center, Kyoto University, Tokai University, The University of Tokushima, KEK, Japan} \\
$^{21}$ {Inst. for Nucl. Research and Nucl. Energy, BG-1784 Sofia, Bulgaria} \\
$^{22}$ {Universit\`a di Pisa, and INFN Pisa, I-56126 Pisa, Italy} \\
$^{23}$ {ICREA and Institute for Space Sciences (CSIC/IEEC), E-08193 Barcelona, Spain} \\
$^{24}$ {Brasileiro de Pesquisas F\'isicas (CBPF/MCTI), R. Dr. Xavier Sigaud, 150 - Urca, Rio de Janeiro - RJ, 22290-180, Brazil} \\
$^{25}$ {NASA Goddard Space Flight Center, Greenbelt, MD 20771, USA and Department of Physics and Department of Astronomy, University of Maryland, College Park, MD 20742, USA} \\
$^{26}$ {Humboldt University of Berlin, Institut f\"ur Physik Newtonstr. 15, 12489 Berlin, Germany} \\
$^{27}$ {Ecole polytechnique f\'ed\'erale de Lausanne (EPFL), Lausanne, Switzerland} \\
$^{28}$ {Finnish Centre for Astronomy with ESO (FINCA), Turku, Finland} \\
$^{29}$ {INAF -- Osservatorio Astronomico di Trieste, Trieste, Italy} \\
$^{30}$ {ISDC - Science Data Center for Astrophysics, 1290, Versoix (Geneva), Switzerland} \\
$^{31}$ {Dip. Di Fisica e Astronomia, Universit\`a degli Studi di Bologna, I-40127 Bologna, Italy} \\
$^{32}$ {INAF -- Istituto di Radioastronomia, I-40129 Bologna, Italy} \\
$^{33}$ {Tuorla Observatory, Department of Physics and Astronomy, University of Turku, Finland} \\
$^{34}$ {Aalto University Mets\"ahovi Radio Observatory, Mets\"ahovintie 114, FI-02540 Kylm\"al\"a, Finland} \\
$^{35}$ {Cahill Center for Astronomy and Astrophysics, California Institute of Technology, Pasadena, CA 91125, USA} \\
$^{36}$ {National Radio Astronomy Observatory (NRAO), PO Box 0, Socorro, NM 87801, USA} \\
$^{37}$ {INAF -- Osservatorio Astrofisico di Torino, I-10025 Pino Torinese (TO), Italy} \\
$^{38}$ {Department of Physics, Purdue University, Northwestern Avenue 525, West Lafayette, IN 47907, USA} \\
$^{39}$ {INAF -- Osservatorio Astronomico di Roma, I-00040 Monteporzio Catone, Italy} \\
$^{40}$ {ASI Science Data Center (ASDC), I-00133 Roma, Italy} \\

\bsp	
\label{lastpage}
\end{document}